\documentclass[preprint]{aastex}
\usepackage{epsf}
\usepackage{natbib}
\usepackage{graphicx}
\usepackage{float}
\usepackage{afterpage}
 \usepackage{relsize}
\usepackage{amsmath}
\usepackage{epsfig}
\usepackage{enumitem}
\usepackage{bm,bbm,amssymb,color, amsmath, color}
\usepackage{enumerate}
\usepackage{url}
\usepackage[utf8x]{inputenc}
\usepackage[section]{placeins}

\usepackage{hyperref}

\begin{document}
\title{MODELING THE NEARLY ISOTROPIC COMET POPULATION IN ANTICIPATION OF LSST OBSERVATIONS}
\author{Kedron Silsbee \altaffilmark{1} \& Scott Tremaine \altaffilmark{2}}
\altaffiltext{1}{Department of Astrophysical Sciences, 
Princeton University, Ivy Lane, Princeton, NJ 08544; 
ksilsbee@astro.princeton.edu}
\altaffiltext{2}{Institute for Advanced Study, 
1 Einstein Drive Princeton, NJ 08540; 
tremaine@ias.edu}
\begin{abstract}
We run simulations to determine the expected distribution of orbital elements of nearly isotropic comets (NICs) in the outer solar system, assuming that these comets originate in the Oort Cloud at thousands of AU and are perturbed into the planetary region by the Galactic tide.  We show that the Large Synoptic Survey Telescope (LSST) should detect and characterize the orbits of hundreds to thousands of NICs with perihelion distance outside 5 AU.  Observing NICs in the outer solar system is our only way of directly detecting comets from the inner Oort Cloud, as these comets are dynamically excluded from the inner solar system by the giant planets.  Thus the distribution of orbital elements constrains the spatial distribution of comets in the Oort cloud and the environment in which the solar system formed.  Additionally, comet orbits can be characterized more precisely when they are seen far from the Sun as they have not been affected by non-gravitational forces.
  \end{abstract}
 
\section{INTRODUCTION}
 The distribution of cometary orbital elements in the Oort cloud depends on the dynamical evolution of the solar system's planetesimal disk and the environment in which the solar system formed.  Unfortunately, the vast majority of Oort cloud comets are unobservable, usually being seen only when they are perturbed onto orbits with perihelion $\lesssim 5$ AU.  Furthermore, the orbits of visible comets have generally been modified by poorly understood non-gravitational forces \citep{Yeomans04}.  For both reasons it is difficult to infer the properties of the Oort cloud from the statistical distribution of comet orbits.
\par 
This paper describes the expected distribution of orbital elements of nearly isotropic comets (NICs). We define these to be comets that have been perturbed into the planetary region from the Oort cloud.  Theoretical models of the distribution of NICs have been constructed by others e.g., \citet{Wiegert99}, \citet{Levison01} and \citet{Fouchard1, Fouchard2, Fouchard3}; the novel feature of the present study is that we focus on much larger heliocentric distances (up to 45 AU) in anticipation of deep wide-angle sky surveys that are currently under development.  We use a simple physical model that assumes a static spherical Oort cloud with comets uniformly distributed on a surface of constant energy in phase space.  Models of the formation of the Oort cloud \citep[e.g.,][]{Dones04} suggest that this approximation is reasonable except perhaps at the smallest semi-major axis we examine, $a=5,\!000$ AU, where the cloud is somewhat flattened.  Furthermore, we assume that these comets evolve solely under the influence of the Galactic tide and perturbations from the giant planets.  We do not consider the stochastic effects of passing stars, as they have effects qualitatively similar to the effects of the Galactic tide \citep{Heisler87, Collins10}; see \citet{Fouchard1, Fouchard2, Fouchard3} for a detailed comparison of the influence of these two agents on the Oort Cloud.  We follow the evolution of these comets through N-body simulations for up to 4.5 Gyr and use the results of these orbit integrations to construct a simulated comet catalog. 
\par
This topic is of special interest now as the Large Synoptic Survey Telescope (LSST) will likely see many of these NICs in the outer solar system.  LSST will survey 20,000 square degrees of sky (48\% of the sphere) about 2,000 times over 10 years, down to an $r$-band magnitude of 24.5 \citep{LSST09}.   LSST has a flux limit 3.2 magnitudes fainter than, and more than three times the area of, the current leader in finding faint distant solar system objects --- the Palomar Distant Solar System Survey \citep{Schwamb10}.  It is expected to find tens of thousands of trans-Neptunian objects \citep{LSST09}; however we are not aware of predictions made specifically for objects originating in the Oort cloud.  
\par
\citet{Francis05} studied the long-period ($P > 200$ years) comet population using the Lincoln Near-Earth Asteroid Research (LINEAR) survey \citep{Stokes00}.  Most observed long-period comets likely originated in the Oort cloud.  He found a sample of 51 long-period comets which were either discovered by LINEAR or would have been, had they not previously been discovered by another group.  Fifteen of these had perihelion distances beyond 5 AU, but none beyond 10 AU.  He used this sample to estimate properties of the Oort cloud.  He found a``suggestive" discrepancy  between the distribution of cometary perihelion distances in the observed sample and in theoretical models \citep{Tsujii92, Wiegert99}, but cautioned that the difference could be the result of a poor understanding of the rate at which comets brighten as they approach the Sun due to cometary activity.  LSST will address this question by observing many comets at large heliocentric radii where they are inactive (see Section \ref{sect:disrupt}).
\par
\citet{Hills81} proposed that the apparent inner edge of the Oort cloud at around 10,000 AU is not due to a lack of comets at smaller semi-major axes, but rather because the perihelion distances of those comets evolve slowly, so they are ejected or evolve to even smaller semi-major axes due to perturbations from the outer planets before they become visible from Earth.  In contrast, comets with semi-major axis $a \gtrsim$ 10,000 AU have their perihelion distance changed by more than the radius of Saturn's orbit in one orbital period, so they are able to jump across the dynamical barrier of the outer planets, and be seen in the inner solar system \citep{Hills81}.  This barrier is not 100\% leak-proof, but as is demonstrated later in the paper, one expects the number density of comets with initial semi-major axes of 10,000 AU to decline by over two orders of magnitude interior to 10 AU.  LSST should detect NICs at distances $>10\,$-15 AU and so will enable us to estimate the population of this inner Oort cloud directly, because we will be able to see NICs outside the region of phase space from which they are excluded by the giant planets.  The properties of this cloud may contain information about the density and mass distribution in the Sun's birth cluster \citep{Brasser12}. 
\par
Observing NICs far from the Sun also probes in unique ways the parts of the Oort cloud that do send comets near Earth.  Non-gravitational forces due to outgassing when the comet comes near the Sun are the primary source of error in determining the original orbits of these comets \citep{Yeomans04}.  It is somewhat uncertain at what radius outgassing begins, but a reasonable estimate would be around 5 AU (see discussion in Section \ref{sect:disrupt}).  Therefore, astrometric observations of comets beyond $\sim \!10$ AU should allow much more precise determination of their original orbits (see discussion at the end of Section \ref{sect:orbel}).
\section{SIMULATION DESCRIPTION}
\label{desc}
We divide phase space into three regions, based on the perihelion distance of the cometary orbit.  We define the ``visibility region" as containing orbits with perihelion distance $q$ in the range $0 \; {\rm AU} < q < 45$ AU.  A ``buffer region" includes orbits with $45 \; {\rm AU} < q < 60\; {\rm AU}$.  All other orbits are defined to be in the ``Oort region". 
 \par
 We simulated orbits with the Rebound package, developed by \citet{Rein12}.  We used their IAS15 integrator, a 15th order adaptive-timestep integrator that is sufficiently accurate to be symplectic to double precision \citep{Rein15}.
 \par
 Our goal is to model the steady-state distribution of NICs with perihelia within 45 AU of the Sun that are produced from an Oort cloud with orbital elements uniformly distributed on an energy surface in phase space (so $dN \sim \sin{I} dI de^2$, where $I$ and $e$ are the cometary inclination and eccentricity).  This approximation assumes that perturbations from the Galactic tide, passing stars or molecular clouds over the last four Gyr have been sufficient to isotropize comets both in position space (seen from the Sun) and velocity space at a fixed position.  This has been shown to be roughly true for comets with semi-major axes greater than 2,000 AU \citep{Duncan87}. 
 \par
 To initialize the simulation, we generated comets at random from this phase-space distribution for four discrete values of initial semi-major axis, $a_i =$ 5,000, 10,000, 20,000, and 50,000 AU, with perihelion distances in the range 60 AU to 60 + $\Delta$ AU.  Then, using an analytic approximation to the torque from the Galactic tide (see appendix \ref{Rtorque}), we determined an upper bound on the time $\tau_{\rm entry}$ (as a function of $a_i$) such that no comet from outside (60 + $\Delta$) AU could enter the buffer or visibility regions within the next $\tau_{\rm entry}$ years.  We chose $\Delta = 5$, but it is straightforward to see that the results do not depend on this choice.  We picked $\tau_{\rm entry} = 10^7, 2.5 \cdot 10^6, 6.25 \cdot 10^5$ and $10^5$ years, for $a_i$ = 5,000, 10,000, 20,000 and 50,000 AU respectively.  These numbers satisfy our upper bound.  
 \par
We then evolved the comets under the influence of the Sun, the Galactic tide, and the four outer planets for $\tau_{\rm entry}$.  After $\tau_{\rm entry}$ had elapsed, we removed any comet with perihelion greater than 60 AU from the simulation.  The remaining comets were allowed to evolve under the influence of the Galactic tide and gravity from the four giant planets and the Sun.  At fixed intervals $\tau_{\rm sample}$ (taken to be 10 years), we recorded the position and velocity of any comet that was within 45 AU of the Sun in a catalog.  This procedure gives us the same expected comet count and distribution of orbital elements as if we had allowed the system to evolve to steady state, and then catalogued the comets visible within 45 AU at an instant in time, and multiplied the flux by $\tau_{\rm entry}/\tau_{\rm sample}$.
\par
Comets are removed from the simulation if they collide with a planet, come within 0.1 AU of the Sun, move outside 200,000 AU, or are perturbed back into the Oort region ($q >$ 60 AU)\footnote{Because the boundary between the buffer and Oort region at 60 AU corresponds to a perihelion distance twice the semi-major axis of Neptune's orbit, we expect the planets to have a negligible effect on the orbits of comets in the Oort region.  Therefore, it is reasonable to assume that a comet with an orbit aligned such that the Galactic tide pulls it from the buffer region into the Oort region will not return for a long time. }.  
\subsection{Orbital Elements}
\label{sect:orbel}
The treatment of orbital elements for highly eccentric orbits that pass through the orbits of massive planets is somewhat subtle.  A comet that is having a close encounter with one of the giant planets will undergo large short-term perturbations to its orbital elements that do not reflect changes to its orbit that will last longer than the duration of the encounter.  Short-term perturbations from such encounters are more serious for comets with large semi-major axes because the energy of the comet in the planetary potential well can equal or exceed the total orbital binding energy of the comet.  At distance $r_p$ from a planet with mass $M_p$, the fractional change in energy due to the potential energy of the planet is 
\begin{equation}
\label{perturbation}
\frac{\Delta E}{E} = \frac{2a_c M_p}{r_p M_\odot} = 0.19\; \frac{a_c}{100 \; {\rm AU}} \frac{1 \,{\rm AU}}{r_p} \frac{M_p}{M_{\rm Jupiter}},
\end{equation}
where $a_c$ is the semi-major axis of the comet prior to the close encounter.  We stress that $a_c$ and $a_i$ are not the same quantity: $a_c$ is the current semi-major axis of the comet, whereas $a_i$ is the semi-major axis of the comet when the simulation was initialized. 
\par
To lessen the short-term planetary perturbations to cometary orbital elements, we adopt the following procedure.  For comets with $a_c < 100$ AU, we simply report heliocentric orbital elements.  These comets have large enough binding energy that they would have to pass close to a planet (within 2.0 AU for Jupiter) to obtain enough extra kinetic energy for $a_c$ to vary by more than 10\% during the close passage (see Equation \eqref{perturbation}).  In order to prevent very close planetary approaches from contaminating our results, we discarded any observation in which the comet is currently close enough to a planet that the specific potential energy due to the planet is more than 10\% of the specific binding energy in an orbit around the Sun with semi-major axis of 100 AU.  This occurs for only $0.004\%$ of all catalog entries, or $0.3\%$ of all catalog entries with $R<10$ AU.
\par
Comets in the visibility region with $a_c > 100$ AU often have potential energies due to the planets which are comparable to their binding energies.  For this reason, we report the barycentric orbital elements of the comet the last time it was in the range [90 AU, 110 AU].  These elements are well-behaved, since they are calculated far outside the orbits of the giant planets where it is appropriate to represent the solar system as having all its mass located at the barycenter.  
\par
The ease with which LSST can determine orbital elements for slowly moving nearly unbound objects is also of interest to this study.  To address this question, we searched the JPL small body database\footnote{\url{http://ssd.jpl.nasa.gov/?horizons}} for objects with semi-major axis greater than 300 AU and perihelion distance greater than 10 AU.  It listed seven objects with a data-arc longer than 5 years.  The estimated errors in $x = 1/a$ ranged from $1.5\cdot 10^{-6}$ AU$^{-1}$ to $1.5 \cdot 10^{-5}$ AU$^{-1}$ for these objects.  It therefore seems reasonable to expect orbits to generally be determined to at least this level of accuracy purely from 10 years of LSST data.
\subsection{Disrupted Comets}
\label{sect:disrupt}
There is a substantial body of evidence suggesting that comets ``fade" over time (e.g., \citealt{Fernandez81, Wiegert99}).  A number of processes have been proposed to explain this phenomenon \citep{Weissman80}: a comet could run out of volatile material, it could have its surface covered with a crust that prevents volatiles from escaping, or it could physically be broken apart by outgassing or tidal stress.  
\par
\citet{Fernandez05} gives 3 AU as a likely cut-off to comet activity based on the sublimation temperature of water, but cautions that many comets experience some activity outside 3 AU due to the sublimation of more volatile elements.   Comet 67P/Churyumov-Gerasimenko first showed signs of activity when it was 4.3 AU from the Sun \citep{Snodgrass13}.  Comet Hale-Bopp showed substantial activity on its approach to perihelion as far out as 7.2 AU \citep{Weaver97}, and at 27.2 AU after perihelion passage \citep{Kramer14}.  When we calculate numbers of visible NICs, we restrict ourselves to NICs further than 5 AU from the Sun.  For this reason, we do not consider the effect of comet activity on magnitude, and just calculate the magnitude from the size and albedo of the bare nucleus, see Section \ref{sect:vis}.  We believe that our assumption that there is negligible activity beyond 5 AU is reasonable, though not certain, given existing observations.  In any event this assumption gives us a conservative estimate of the number of comets that a survey like LSST will discover.
\par
Because comet activity does not affect brightness in our model, we are only sensitive to physical disruption of comets, not loss of volatiles.  For this reason, throughout this paper, we refer to ``disruption", rather than ``fading".

In the results that follow, we remove a comet after it has made 10 apparitions in the catalog with radius $R < 3$ AU (corresponding to a total exposure to the Sun at $R < 3$ AU of about 100 years, since $\tau_{\rm sample} = 10$ years).  Comets are also removed if they ever travel within 0.1 AU of the Sun (even if they do not appear so close in the catalogue) or if they suffer a collision with one of the gas giants.
\section{COMET LIFETIMES IN THE SIMULATION REGION}
\label{lifetime}
\begin{figure}
\centering
\includegraphics[scale = .32]{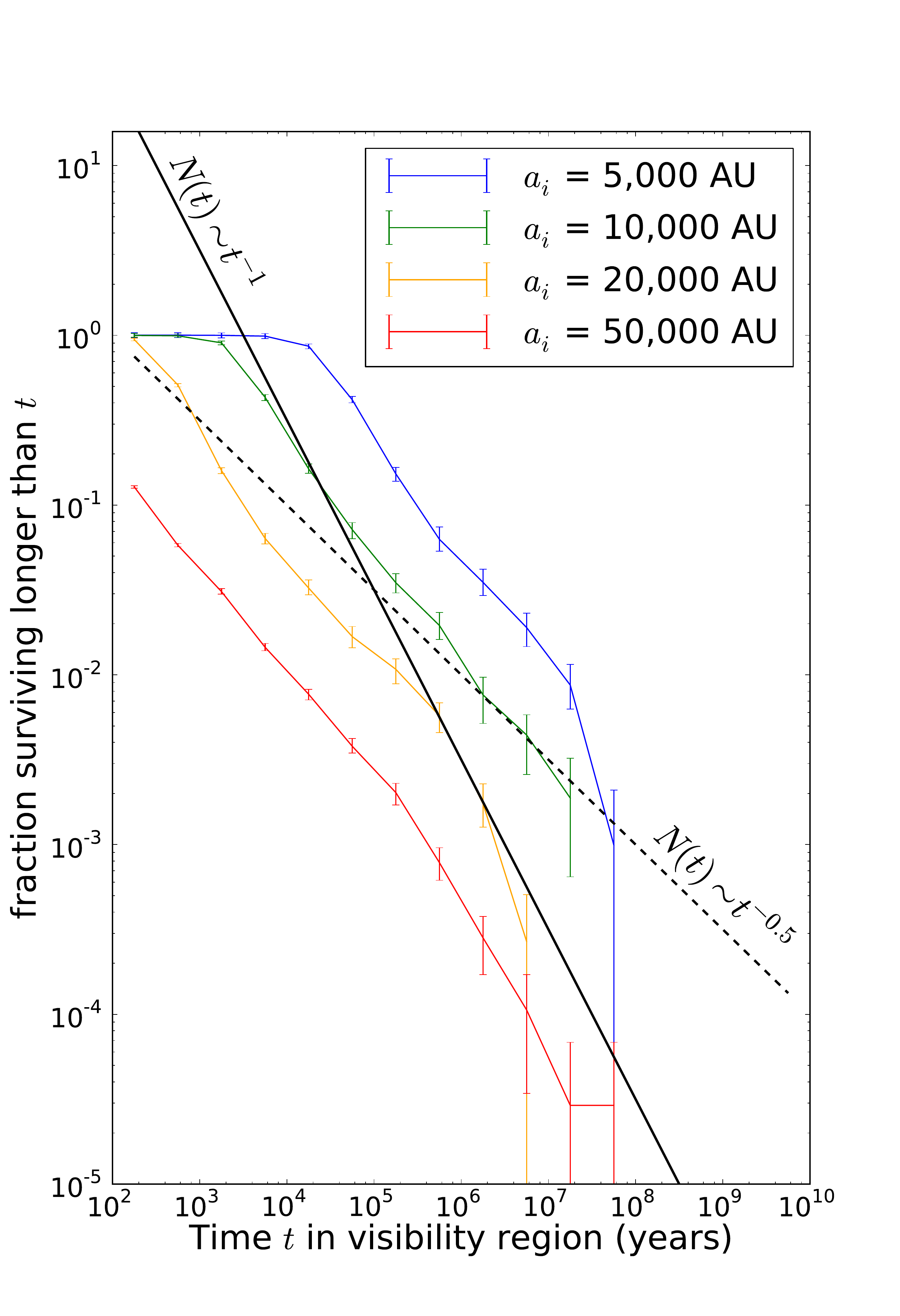}
\caption{Distribution of the number of NICs having $R<45$ AU for longer than $t$ years.  The curves show four different values of the initial semi-major axis $a_i$, normalized by the number of comets in the simulation which ever enter within $R = 45$ AU.   We terminate the integrations after 4.5 Gyr. The black lines show two power laws with exponents $-0.5$ and $-1$.  Only comets with $a_c > 34.2$ AU (corresponding to long-period comets --- comets with period $>$ 200 years) were counted.}
\label{Occurrence}
\vspace{-.05cm}
\end{figure}

Figure \ref{Occurrence} shows the fraction of NICs in our simulations with $a_c > 34.2$ AU (period $>$ 200 years) appearing in the region with $R<45$ AU for more than $t$ years, as a function of $t$.  Different curves correspond to different values of the initial semi major axis $a_i$.  In this plot we terminate each orbit integration after 4.5 Gyr.  The error bars are derived from the re-sampling procedure described in Section \ref{sect:concentration}.  In this and all subsequent plots, only comets with periods greater than 200 years (corresponding to semi-major axes greater than 34.2 AU) are counted.
\par
\citet{Yabushita79} argued using a simple random walk model that the number of NICs surviving more than $N_{\rm peri}$ perihelion passages should scale as $P(>\! \!N_{\rm peri}) \propto N_{\rm peri}^{-1/2}$.  Assuming for the sake of argument that NICs spend a fixed amount of time in the visibility region per perihelion passage, then the number of NICs having a given number of catalog entries is proportional to the number of NICs surviving for more than a given number of perihelion passages.  We should therefore recover the same power-law as \citet{Yabushita79} if his model is a good approximation to the full physics captured by the simulation.  This plot largely confirms the predictions of \citet{Yabushita79}, but the exponent of the power-law seems to be slightly steeper than his value of $-1/2$.  
  \par
 Deviations from power-law behavior at short times occur because the visibility region is larger than the region of influence of the planets, so there is some delay before NICs that have entered the visibility region interact with the planets.  Comets with smaller values of $a_i$ experience less torque due to the Galactic tide, so the delay is larger.  They also have more binding energy that must be overcome prior to ejection.  This explains the trend seen in Figure \ref{Occurrence} that NICs with smaller $a_i$ take longer to be ejected.
\par
The fact that some of our simulated particles survive for longer than 1 Gyr leads to concern about the physical validity of our assumption that there is a static Oort cloud.  Likely, many of the NICs that will be observed with LSST exited the Oort cloud more than 1 Gyr in the past, when it may have had different physical properties.
\par 
Even if the properties of the Oort cloud have not changed over 4.5 Gyr, long-lived comets may bias our simulations, as the following argument shows.  Suppose, as seems reasonable from Figure \ref{Occurrence}, that the true distribution of time $t$ that a comet spends in the visibility region is given by a power law, i.e.
\begin{equation}
\label{simpledndt}
dp/dt = \left\{
\begin{array}{ll}
\frac{(\alpha-1) t^{-\alpha}}{t_{\rm min}^{1-\alpha}}  & t > t_{\rm min} \\
0 & t<t_{\rm min}
\end{array}
\right.
\end{equation}
for some $\alpha$ in the interval $1<\alpha < 2$.  Then, if we terminate our integrations at some time $t_{\rm cutoff}$, the average time that a comet spends in the visibility region during our simulation is
\begin{equation}
\langle t \rangle = \frac{1}{2-\alpha} \left[t_{\rm cutoff} \left(\frac{t_{\rm min}}{t_{\rm cutoff}}\right)^{\alpha-1} + (1-\alpha) t_{\rm min}\right]
\end{equation}
In the limit that $t_{\rm cutoff} \gg t_{\rm min}$, this becomes 
\begin{equation}
\label{expectedt}
\langle t \rangle = \frac{t_{\rm cutoff}}{2-\alpha}\left(\frac{t_{\rm min}}{t_{\rm cutoff}}\right)^{\alpha - 1} 
\end{equation}
\par
In our simulations, (8, 1, 0, 0) particles with $a_i$ = (5,000, 10,000, 20,000, 50,000) AU survive with $q < 45$ AU for the duration of the integration (4.5 Gyr).  For $a_i = 20,\!000$ AU and $50,\!000$ AU respectively, the longest lived particles survived for 1.1 and 1.4 Gyr.  Now, consider the case of the simulation with $a_i = 20,\!000$ AU.  We drew particles with lifetimes from the distribution in Equation \eqref{simpledndt}.  By chance, we drew no particles with lifetime greater than $t_{\rm max} = 1.1$ Gyr.  We will assume that we drew randomly from the distribution in Equation \eqref{simpledndt} subject to the constraint that we draw no particles with lifetimes greater than $t_{\rm max}$ (this is not quite true, given that we did draw one particle with lifetime exactly $t_{\rm max}$).  We can renormalize the distribution in Equation \eqref{simpledndt} subject to the constraint that no comet have lifetime longer than $t_{\rm max}$, and calculate the mean.  This is given by (assuming $t_{\rm max} \gg t_{\rm min}$)
\begin{equation}
\label{truncatedt}
\langle t \rangle = \frac{\alpha-1}{2-\alpha} \left[t_{\rm max} \left(\frac{t_{\rm min}}{t_{\rm max}}\right)^{\alpha - 1}\right]
\end{equation}
Taking the ratio of Equations \eqref{expectedt} to \eqref{truncatedt}, and assuming that a comet makes a fixed number of appearances in the catalog per unit time spent in the visibility region, we find we have likely underestimated our total flux by something close to a factor of $(4.5/1.1)^{2-\alpha}/(\alpha-1)$.  This expression is 4.0 if we take $\alpha = 1.5$, as suggested by \citet{Yabushita79}, but declines to unity for $\alpha = 2.0$.  In principle, this bias can be reduced by simulating more comets but the computational resources necessary to reduce the bias substantially are prohibitive.

\par
In the following sections we present simulation data for different values of $t_{\rm cutoff}$.  $t_{\rm cutoff}$ is defined relative to the time when the comet first enters within 45 AU {\it except} when $t_{\rm cutoff} = 4.5$ Gyr, in which case it is defined relative to the start of the simulation.  We believe that a value of a few Gyr is most observationally relevant, and most of the following discussion is based on such cutoff times, however given the limitations discussed above, we show results for shorter cutoff times as well.
\section{CONCENTRATION OF NICS DUE TO THE GIANT PLANETS}
In this section we describe how the distribution of NICs is affected by the planets.
\subsection{Expected Distribution of $R$ and $q$ with no Planets}
We first calculate the expected distribution of orbital radius $R$ and perihelion distance $q$ in the absence of perturbations from the planets, constructing what we call the zero-planet model.  We assume a uniform distribution of cometary orbital elements in phase space at four fixed energies.  These results provide a natural normalization to the plots in the following subsection, which show the distributions of $R$ and $q$ in our simulated catalog.
\subsubsection{Distribution in Radius and Perihelion at a Snapshot in Time}
Let there be $N_0$ comets distributed uniformly on the energy surface corresponding to semi-major axis $a$.  There is no need to distinguish between the initial semi-major axis $a_i$ and the current semi-major axis $a_c$ here, since the semi-major axis is not changed in the absence of planetary perturbations.  Approximating the orbits as parabolic in the visibility region, we find that the radial velocity at radius $R$ of an orbit with perihelion distance $q$ is
\begin{equation}
\label{vr}
v_R(q, R) = \sqrt{\frac{2GM_\odot(1-q/R)}{R}}.
\end{equation}
Since we assume a uniform distribution of comets on the energy surface, the probability density of the squared eccentricity $e^2$ is 
\begin{equation}
N(e^2) de^2 = N_0de^2.
\end{equation}
Then, since $q=a(1-e)$, the number of comets per unit perihelion distance $N(q)$ is given by 
\begin{equation}
\label{Nq}
N(q)dq = 2N_0\left(1-\frac{q}{a}\right )\frac{dq}{a}\simeq \frac{2N_0}{a} dq,
\end{equation}
where the last equality holds because we are interested in comets with $q \ll a$.  A comet on a near-parabolic orbit with perihelion $q$ spends a fraction of its time $f(R,q)dR$ in the radial interval between $R$ and $R+dR$, where 
\begin{equation}
\label{frq}
f(R, q)dR = \frac{2 dR}{Pv_R(q, R)},
\end{equation}
and $P$ is the period of the orbit.  Then, using Equations \eqref{Nq} and \eqref{frq}, we can solve for the number of comets in a radial interval $N(R)dR$:
\begin{equation}
\label{Nr}
N(R) = \int_{q=0}^R N(q) f(R, q) dq = \frac{2 \sqrt{2} N_0 R^{3/2}}{\pi a^{5/2}}.
\end{equation}
\par
The number of comets with perihelion in the range $q$ to $q + dq$ expected to be present out to a maximum value of $R$ is given by 
\begin{equation}
N(q|R_{\rm max})dq = N(q)dq \int_q^{R_{\rm max}} f(R', q) dR'.
\end{equation}
This evaluates to 
\begin{equation}
\label{Nqexpect}
N(q|R_{\rm max})dq = \frac{2 \sqrt{2}}{3 \pi a^{5/2}} \left(\frac{R_{\rm max}}{q}-1\right)^{1/2} \left(2 + \frac{R_{\rm max}}{q}\right)q^{3/2} dq.
\end{equation}
As mentioned in Section \ref{desc}, because of the way we have set up our simulation, and the fact that we sample every $\tau_{\rm sample}$ years, we would expect our catalog to contain a number of comets with orbital elements $\psi$ equal to 
\begin{equation}
\label{sampleToActual}
N_{\rm cat}(\psi|R_{\rm max})= \frac{\tau_{\rm entry}}{\tau_{\rm sample}} N(\psi|R_{\rm max}),
\end{equation}
if we had not included any planets in the simulation.
\subsubsection{Concentration Factors}
\label{sect:concentration}
\begin{figure}
\centering
\includegraphics[width=.5\textwidth]{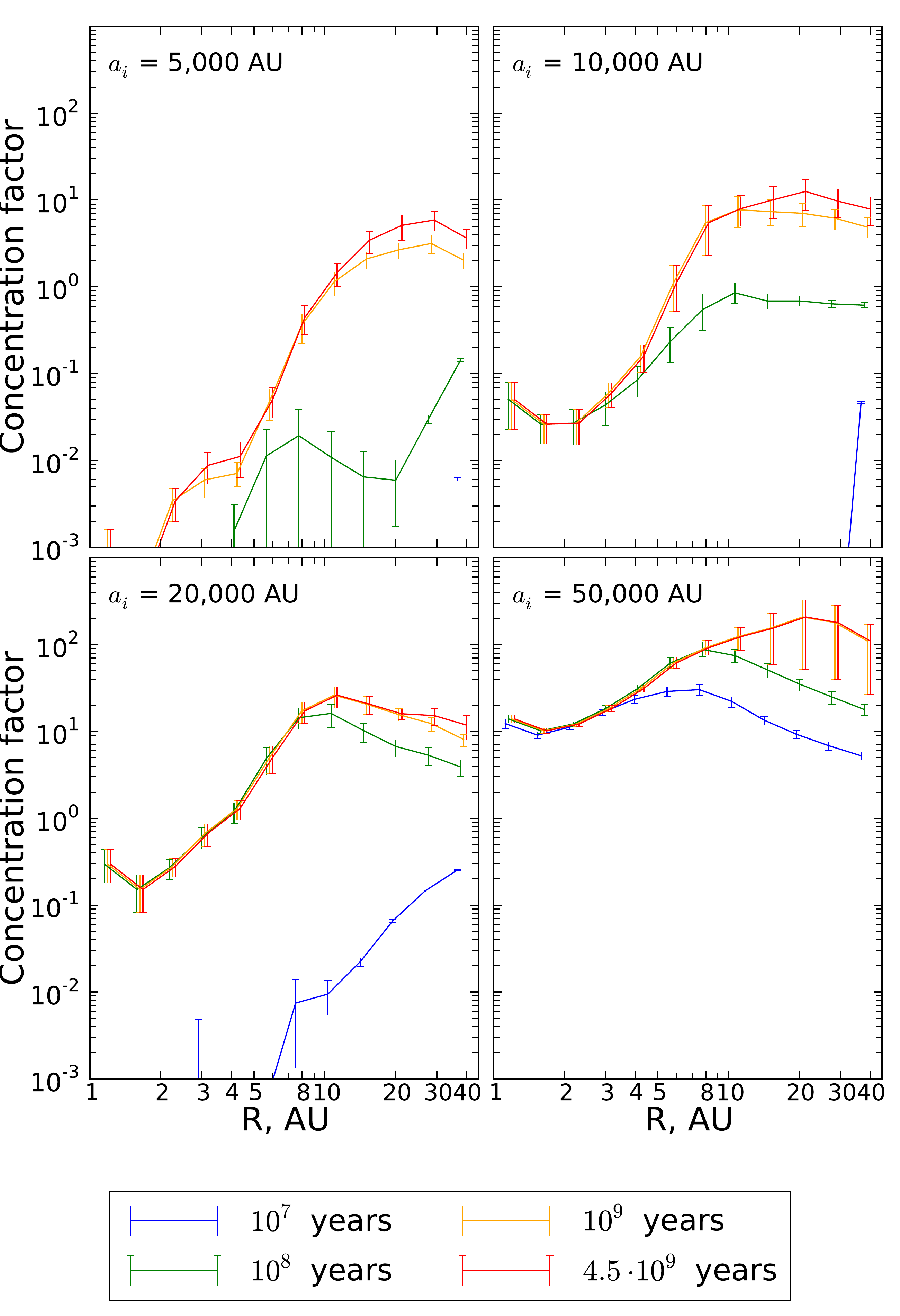}
\caption{Number of comets in the simulated catalog at radius $R$, normalized by the expected number assuming no planets (Equations \eqref{Nr}, \eqref{sampleToActual}).  Different curves correspond to different values of the cutoff time $\tau_{\rm cutoff}$.  Errors were determined via bootstrapping (see Section \ref{sect:concentration} for details).  Points and error bars from curves corresponding to different cutoff times have been horizontally displaced slightly for clarity.  }
\label{Rconcentration}
\vspace{-.05cm}
\end{figure}

In this subsection we compare our catalog to the one which would be produced had we not included planets in our simulations.  In Figure \ref{Rconcentration}, we compare our simulated comet catalog with the zero-planet model (Equations \eqref{Nr} and  \eqref{sampleToActual}).  We have plotted a ``concentration factor" --- the ratio of comets appearing in our catalog within a given radial interval to the number calculated from Equations \eqref{Nr} and \eqref{sampleToActual}, assuming the same density of comets in the Oort cloud as was used to initialize our simulations.  As stated previously, only comets with periods greater than 200 years are counted.  Each panel corresponds to a different value of the initial cometary semi-major axis $a_i$.  Different colored lines correspond to different values of $\tau_{\rm cutoff}$ as shown in the legend. 
\par
These data represent the results from following $N_{\rm sim}$ comets where $N_{\rm sim} =$ 10,000, 16,000, 40,000, and 420,000 for $a_i =$ 5,000, 10,000, and 20,000 AU,  and 50,000 AU respectively.   Note that fewer than 10\% of these ever enter the region $R < 45$ AU (955, 1548, 3949, and 29215 respectively).  Most comets do not evolve to $q < 60$ AU within the entry period and are therefore discarded at the end of the entry period.  Some of those that do come within $q < 60$ AU never reach $q < 45$ AU (if the angular momentum is nearly perpendicular to the torque).
\par 
We would like to know the true distribution of orbital elements in the limit that we simulate a very large number of comets.  To estimate our random error, we employed re-sampling.  For each point on the curve, we drew $N_{\rm sim}$ comets with replacement from our $N_{\rm sim}$ simulated comets.  The points are the mean of the re-sampled distribution, and the error bars correspond to the 16$^{\rm th}$ and 84$^{\rm th}$ percentiles of the re-sampled distribution (if the distribution were normally distributed, these would be 1-sigma error bars).  The error bars on nearby points are highly correlated in most of our plots because the same comet contributes to several bins in the course of its evolution, hence the low point-to-point scatter relative to the error bars.  
\par
Although these data reflect the orbits of thousands of comets, the statistical errors are large in many cases, since often the majority of the contribution to a particular bin comes from only one or two long-lived comets, (see the discussion in \S \ref{lifetime}).

\begin{figure}
\centering
\includegraphics[width=.5\textwidth]{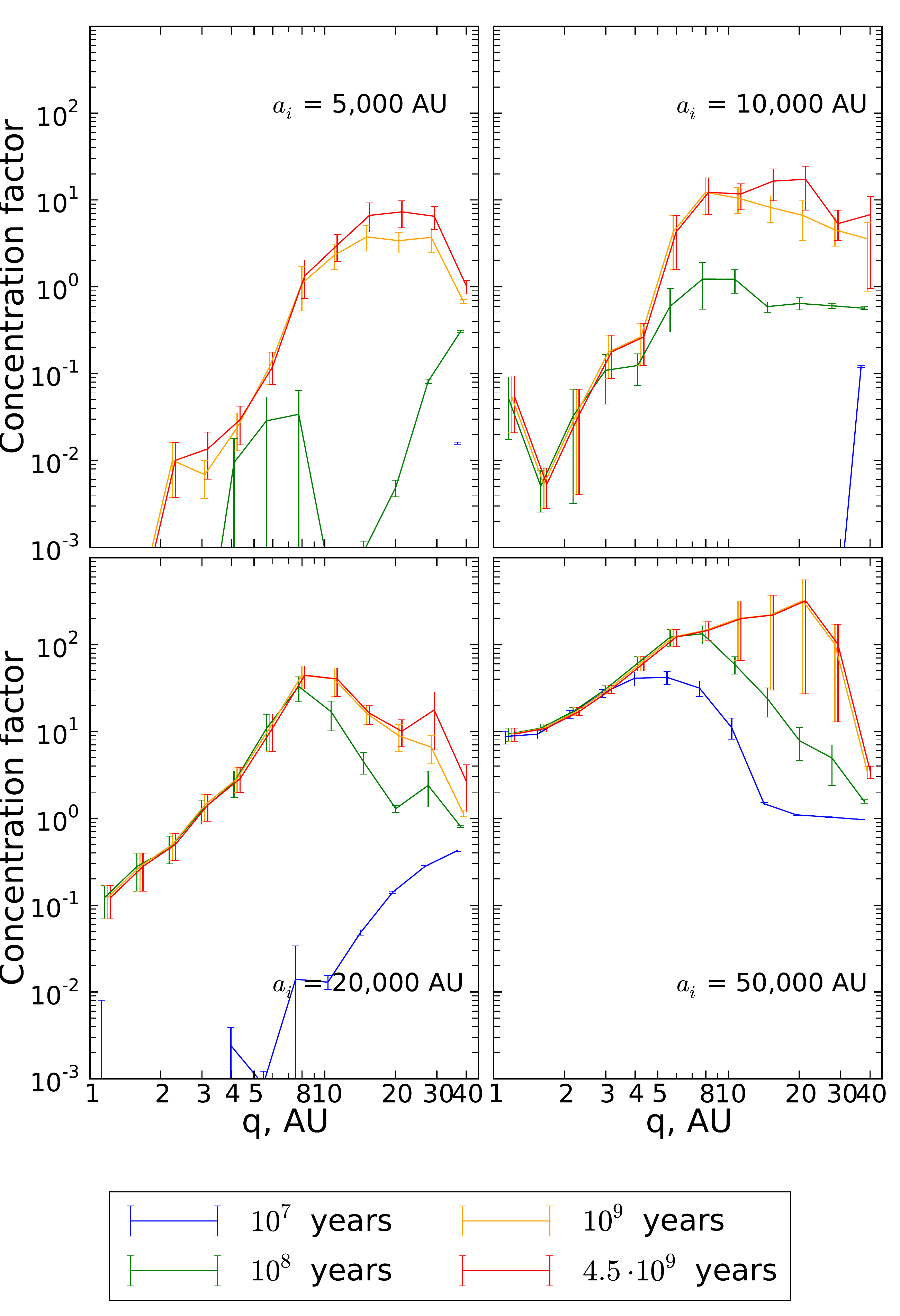}
\caption{Number of comets in the catalog at perihelion $q$, normalized by the expected number assuming no planets (Equations \eqref{Nqexpect}, \eqref{sampleToActual}).  Different curves correspond to different values of the cutoff time $\tau_{\rm cutoff}$.}
\label{Qconcentration}
\vspace{-.05cm}
\end{figure} 
Figure \ref{Qconcentration} is similar to Figure \ref{Rconcentration}, except that instead of plotting the number of appearances in our catalog in bins of heliocentric radius $R$, we plot appearances in bins of perihelion $q$.  The distribution of $q$ provides more information about the NIC orbits: a sharp feature in the perihelion distribution is smoothed out when one looks at the distribution of heliocentric radius, since the same orbit can be observed at a range of values of $R$.   
 \par
In Figures \ref{Rconcentration} and \ref{Qconcentration}, and in most of the subsequent plots, we have horizontally offset the blue, green, yellow and red curves  curves by $-4.5\%$, $-1.5\%$, $1.5\%$, and $4.5\%$ respectively in order to make the curves distinguishable in regions where they overlap.
\par
The following qualitative features of Figures \ref{Rconcentration} and \ref{Qconcentration} are straightforward to explain:
\begin{itemize}
\item The flux of comets with $a_i \lesssim 20,\!000$ AU shows a sharp drop-off (in both $N(q)$ and $N(R)$)  interior to 10 AU.  This is because comets originating at small semi-major axes are subjected to weak Galactic tides and change their perihelia slowly.  The majority of kicks given to a comet with $q<10$ AU are large enough to either unbind a comet infalling from outside a few thousand AU, or to reduce $a_c$ to the point that the timescale to change the perihelion distance is much longer than an orbital period.   We therefore expect to see a jump in the number of comets appearing inside 5 AU at values of $a_i$ exceeding that at which a comet can go from perihelion $\gtrsim 15$ AU (largely unaffected by Jupiter and Saturn) to perihelion $\lesssim 5$ AU in one orbit.  This occurs at approximately 30,000 AU.  Therefore, in order for a comet starting from inside $\sim 30,\!000$ AU to appear in the inner solar system, it needs to have a lucky orientation with respect to Jupiter and Saturn\footnote{Or a kick from a star that passes unusually close to the Sun, a rare event not included in our model.}.  This lucky orientation can either yield multiple small energy kicks on subsequent perihelion passages, or yield a kick that increases the semi-major axis so that the comet receives a larger torque from the Galaxy \citep{KaibQuinn09}.  
 \par   
 Thus we conclude that comets with $a_i \lesssim 20,\!000$ AU are mostly ejected by interactions with the outer planets before they reach small heliocentric radii.  \citet{Hills81} arrives at a similar result considering the effects of passing stars discretely rather than as a smooth Galactic tidal potential.  \citet{Collins10} provide a discussion of when it is appropriate to treat the influence of the Galaxy as being due to a smooth tidal field, and when it should be modeled as discrete stellar encounters.  
 \par
 The lower concentration factors for comets with smaller values of $a_i$ do not mean that we will see fewer NICs for a given number of Oort cloud comets at that energy.  This is because the fraction of the total comets that are in the visibility region at a given time in the zero-planet model scales with $a^{-2.5}$ (see Equation \eqref{Nr}). 
\item 
The difference between the green curves and the orange and red curves grows with increasing $q$.  This is because the kicks from the planets are smaller at large $q$, so the comets survive longer.  The exception to this is the curve for $\tau_{\rm cutoff} = 100$ Myr and $a_i =$ 5,000 AU.  In this case, comets have mostly had insufficient time to be torqued to small values of $q$.  It should be noted however, that the time to reach a given perihelion distance is not completely determined by the initial semi-major axis because a comet could be scattered to larger $a_i$ by an early encounter with Neptune, and subsequently evolve more rapidly.  
\par
There is little difference between the results for $t_{\rm cutoff} = 1$ Gyr, and $t_{\rm cutoff} = 4.5$ Gyr for comets with $a_i \geq 20,\!000$ AU.  As discussed previously, this is likely an artifact of our simulations including insufficiently many comets with these values of $a_i$ to capture the tail of the lifetime distribution (see Section \ref{lifetime}).
\item
The concentration factors for large cutoff time in Figure \ref{Qconcentration} approach unity as $q$ approaches 45 AU, however the concentration factors in Figure \ref{Rconcentration} are still on the order of 10 at 45 AU.  This is because even a concentration of comets with $q \ll 45$ AU affects the distribution of comets at $R=45$ AU. 
\item
We do not expect $N(q)$ to drop exactly to unity as soon as $q$ is larger than the extent of the planetary perturbations, because NICs which have interacted with the planets may be systematically carried away from the planetary region by the tide at a different rate than they were carried in (due to a change in orientation or semi-major axis). 
\end{itemize}
\section{DISTRIBUTION OF ORBITAL ELEMENTS FOR VISIBLE COMETS}
The above analysis shows the degree to which the giant planets concentrate NICs in the outer solar system and exclude them from inside the orbit of Jupiter.  In this section, we use an estimate of the size distribution of NICs, the relationship between magnitude, size and heliocentric distance, and the concentration effect due to interactions with the planets to calculate the number of NICs expected to be seen in an all-sky survey as a function of the limiting magnitude $m_{\rm lim}$.   
\subsection{Size Distribution}
Comets have so far been observed primarily within a few AU of the Sun, where their brightness is influenced strongly by their activity.  At the larger distances that we focus on here, comets are believed to be generally inactive (see discussion in Section \ref{sect:disrupt}), so their brightness is determined solely by their size, distance, and albedo.  Let $H$ be the apparent magnitude of a comet 1 AU from the Sun and 1 AU from the observer, seen from zero phase angle.  Based on a sample of long-period comets from about $H = 5$ to $H = 9$, \citet{Sosa11} derive a relation for active comets between the radius $r$ (in kilometers) and $H$:
\begin{equation}
\label{rH}
\ln{r} = \alpha+\beta H,
\end{equation}
where $\alpha$ = 2.072 and $\beta = -0.2993$.  We caution that our use of this formula requires substantial extrapolation: the largest comet used to determine the formula has a radius of 1.8 km, more than an order of magnitude smaller than the smallest comets detectable at 30 AU in a survey with the limiting magnitude of LSST (see Section \ref{sect:vis}).  \citet{Sosa11} note that the relation in Equation \eqref{rH} predicts a radius (13 km) for comet Hale-Bopp that is somewhat below other estimates (mostly falling in the 20--35 km range).  This discrepancy suggests that Equation \eqref{rH} may underpredict the radii of large comets, in which case our estimates of the observable comet population will be conservative.  Note that by using Equation \eqref{rH} we are assuming that long-period comets are mostly the same population as NICs (or at least have the same size distribution).
\par
\citet{Hughes01} finds that the number $N_{\rm peri}$ of long-period comets with brightness $H < 6.5$ passing through perihelion in the inner solar system per year per AU of perihelion distance is given by 
\begin{equation}
\frac{dN_{\rm peri}}{dH} = c_0 e^{\gamma H},
\end{equation}
with $c_0 = 2.047 \cdot 10^{-3}$ and $\gamma =  0.827$.  We can then transform variables to $r$ using Equation \eqref{rH}.  We find that
\begin{equation}
\label{dNdr}
dN_{\rm peri}/dr =  -\frac{c_0}{\beta} \mathlarger{\mathlarger{e^{-\gamma \alpha/\beta} r^{\gamma/\beta - 1}}} = 2.09\cdot r^{-3.76}.
\end{equation}
\par
This distribution holds down to $r(H = 6.5) = 1.1$ km, however, for simplicity we extrapolate it down to $0.9$ km --- the smallest comet visible at 5 AU in our model (see next section).  This size distribution leads to a weak divergence in total mass at the large end of the spectrum.  Nevertheless, we assume that the power-law behavior holds up to several tens of kilometers.  The size distribution in Equation \eqref{dNdr} is steeper than the relation ($dN_{\rm peri}/dr \sim r^{-2.79}$) estimated in \citet{Hughes01} because he uses a different relation between $H$ and $r$.  It is also substantially steeper than the relation ($dN_{\rm peri}/dr \sim r^{-2.92}$) found in \citet{Snodgrass11} for the Jupiter-family comets.  If the size distribution is shallower than we have estimated at large radii, then our estimates of the observable comet population will be conservative. 
\subsection{Visibility Model}
\label{sect:vis}
In this section we describe our model for determining how likely a given simulated comet is to be visible.  We assume that comets have an $r$-band geometric albedo $A_g$ of 0.04 as suggested in \citet{Lamy04}.  We find that the magnitude $m$ of an inactive comet is 
\begin{align}
\label{mag}
m &= -27.08 - 2.5 \log{\left(\frac{r^2 A_g {\rm (AU)}^2}{R^4}\right)} 
\nonumber\\
&= 24.28 - 2.5 \log{(A_g/0.04}) -5 \log{(r/{\rm 1km})} + 10 \log{(R/{\rm 5AU})},
\end{align}
where we have used $-27.08$ as the apparent $r$-band magnitude of the Sun.  Equation \eqref{mag} is only valid for comets far from the Sun, since we have assumed that $R_{\rm Sun, comet} = R_{\rm Earth, comet}$, that the phase angle is zero, and most importantly, that we are seeing the bare nucleus of an inactive comet.  \citet{Lamy04} state that magnitude drops off at about a rate of 0.04 magnitudes/degree of phase angle, meaning that error due to this nonzero phase angle is limited to at most 0.23 magnitudes for a comet at 10 AU.  Similarly, at 10 AU, the largest possible error arising from the approximation that the Sun-comet distance is equal to the Earth-comet distance also corresponds to a magnitude error of $\Delta m = 0.23$.
\subsection{Weighting of Observations}
Using Equation \eqref{mag} we can solve for $r_{\rm min}(R)$, the radius in kilometers of the smallest comet visible at distance $R$ in a survey with limiting magnitude $m_{\rm lim}$, assuming $A_g = 0.04$:
\begin{equation}
\label{rR}
r_{\rm min}(R)  = 0.903 \cdot 10^{0.2(24.5-m_{\rm lim})}  \left(\frac{R}{\rm 5\; AU}\right)^2.
\end{equation}
The comet size is not explicitly tracked in our simulations.  We assume that the sizes of comets in our simulation are drawn from the distribution in Equation \eqref{dNdr}, with a lower cutoff radius of $0.903$ km --- the smallest comet visible at 5 AU in our model.  To account for the fact that not all comets are visible at all orbital radii, we weight a simulated comet appearance at radius $R$ by the fraction of comets that would be visible at the observed value of $R$ given the assumed size distribution in the simulation.  An appearance at high $R$ will receive a low weight, since most comets would not be visible so far away.  We assign weight 1 to observations at $R = 5$ AU, as all comets in our assumed size distribution would be visible at 5 AU.  Then at general $R$, we assign weight 
\begin{equation}
W(R) = \frac{\int_{r_{\rm min}(R)}^\infty r^{-3.76} dr}{\int_{r_{\rm min}(5 \; {\rm AU})}^\infty r^{-3.76} dr} = \left(\frac{5 \; {\rm AU}}{R}\right)^{5.52}.
\end{equation}
\par 
We quote numbers of comets with a given set of orbital elements per $10^{11}$ comets larger than one kilometer in a  spherical distribution at the assumed initial semi-major axis.  To achieve this normalization, we multiply our counts by
\begin{equation}
\frac{10^{11}f_{60-65}(a_i)}{N_{\rm init}} \frac{\tau_{\rm sample}}{\tau_{\rm entry}} \frac{\int_{0.903}^\infty N(r) dr}{\int_1^\infty N(r)dr},
\end{equation}
where $f_{60-65}(a_i)$ is the fraction of the phase space of orbits with semi-major axis $a_i$ that consists of orbits with perihelia between 60 and 65 AU, given by 
\begin{equation}
f_{60-65}(a_i) = \frac{(a_i - 60 \; {\rm AU})^2 -(a_i - 65 \; {\rm AU})^2}{a_i^2},
\end{equation}
and $N_{\rm init}$ is the number of comets we initialize between 60 and 65 AU.

\subsection{Distribution of Visible Comets}
\begin{figure}
\centering
\includegraphics[width=.5\textwidth]{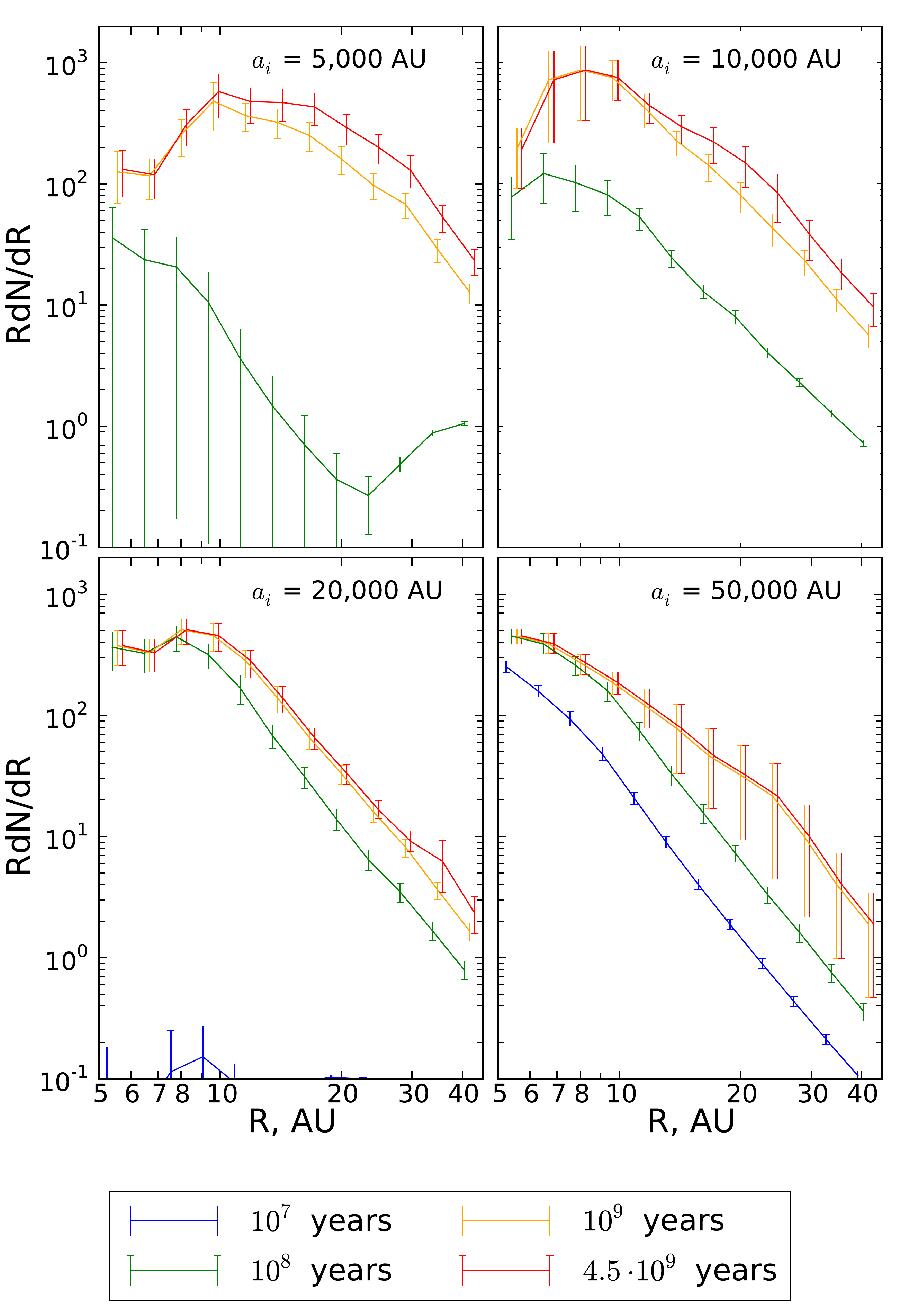}
\caption{Number of NICs expected to be seen per logarithmic interval in $R$ at a snapshot of time in an all-sky survey with limiting $r$-band magnitude $m_{\rm lim} = 24.5$.  This assumes there are $10^{11}$ comets with $r > 1$ km at the value of initial semi-major axis $a_i$ specified in each panel.  Errors were determined via bootstrapping (see Section \ref{sect:concentration} for details).  Different curves correspond to different values of $t_{\rm cutoff}$.}
\label{Rvisible}
\vspace{-.05cm}
\end{figure} 
 \begin{figure}
\centering
\includegraphics[width=0.5\textwidth]{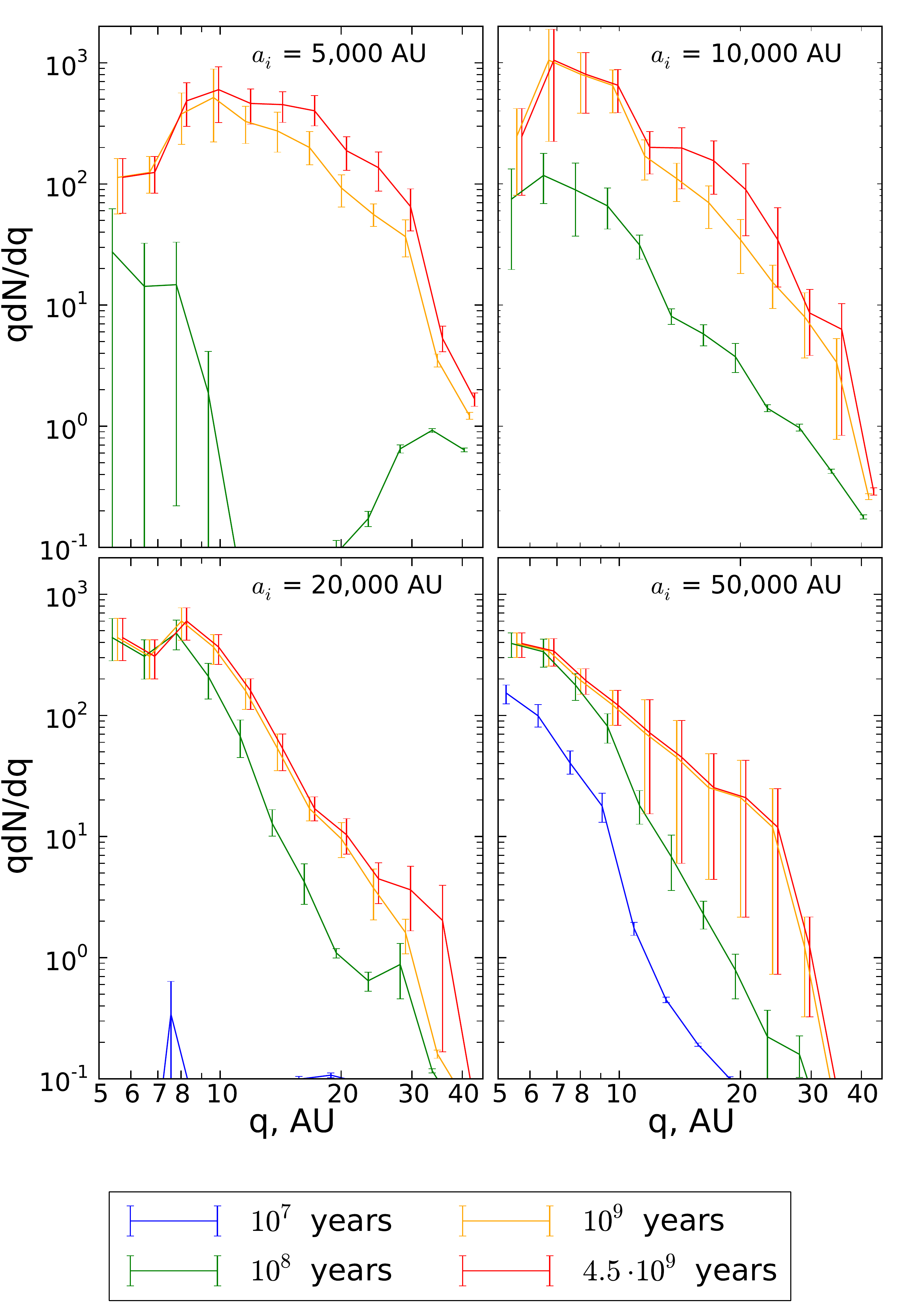}
\caption{Number of NICs expected to be seen per logarithmic interval in $q$ at a snapshot of time.  This assumes there are $10^{11}$ comets with $r > 1$ km at the value of $a_i$ specified in each panel.  Different curves correspond to different values of $t_{\rm cutoff}$.}
\label{Qvisible}
\vspace{-.05cm}
\end{figure} 
\
In this section we present results from our simulations showing how many NICs are visible over the whole sky at a given snapshot of time as a function of $R$ and $q$, for observations taken between $R = 5 \; {\rm AU}$ and $R = 45 \; {\rm AU}$.  In all cases, we assume a limiting $r$-band magnitude $m_{\rm lim}$ of 24.5, equivalent to the one-exposure limit for LSST \citep{LSST09}.  The number of distant NICs expected to be discovered by LSST differs from the results presented here for two reasons.  First, LSST is expected to operate for 10 years, so it should see more than just the comets visible in a snapshot, particularly in the case of the closer comets where $R/v$ is less than 10 years.  Second, LSST will only survey 48\% of the sky, so will only see about half of the comets that would be visible in an equivalent all-sky survey.  Comets move slowly enough that trailing losses will be insignificant given the 30 second exposure time.  Using Equation (8) from \citet{Ivezic08}, we estimate a comet at 10 AU will have a limiting magnitude only 0.06 magnitudes brighter due to trailing losses.
\par
Figures \ref{Rvisible} and \ref{Qvisible} show the number $N$ of NICs expected to be visible outside $5$ AU per unit of $\ln{R}$ and $\ln{q}$ respectively, per $10^{11}$ comets with $r$ greater than 1 km at the labeled initial semi-major axis in the Oort cloud.  The shapes of the curves are substantially different for different values of $a_i$, particularly in the region between 5 and 10 AU, where the statistics are the best.  In the $a_i =$ 5,000 AU case, the expected count {\it increases} by a factor of 5 from $R=$ 5 AU to $R=$ 10 AU for $t_{\rm cutoff} \geq 1$ Gyr.  In the $a_i =$ 50,000 AU case it {\it decreases} by a factor of $\approx 5$.  Observations of comets in this regime will therefore allow us to observationally constrain the distribution of $a_i$.  The peak of $RdN/dR$ moves smoothly from around 15 AU for $a_i = 5,\!000$ AU to less than 5 AU for $a_i = 50,\!000$ AU.
\par
As shown in Appendix \ref{sect:isovis}, we expect $RdN/dR$ and $qdN/dq$ to decline as $R^{-3.02}$ and $q^{-3.02}$ respectively in the zero-planet model.  Deviations from this behavior are due to variation in the concentration factor as shown in Figures \ref{Rconcentration} and \ref{Qconcentration}.
 \begin{figure}
\centering
\includegraphics[width=.5\textwidth]{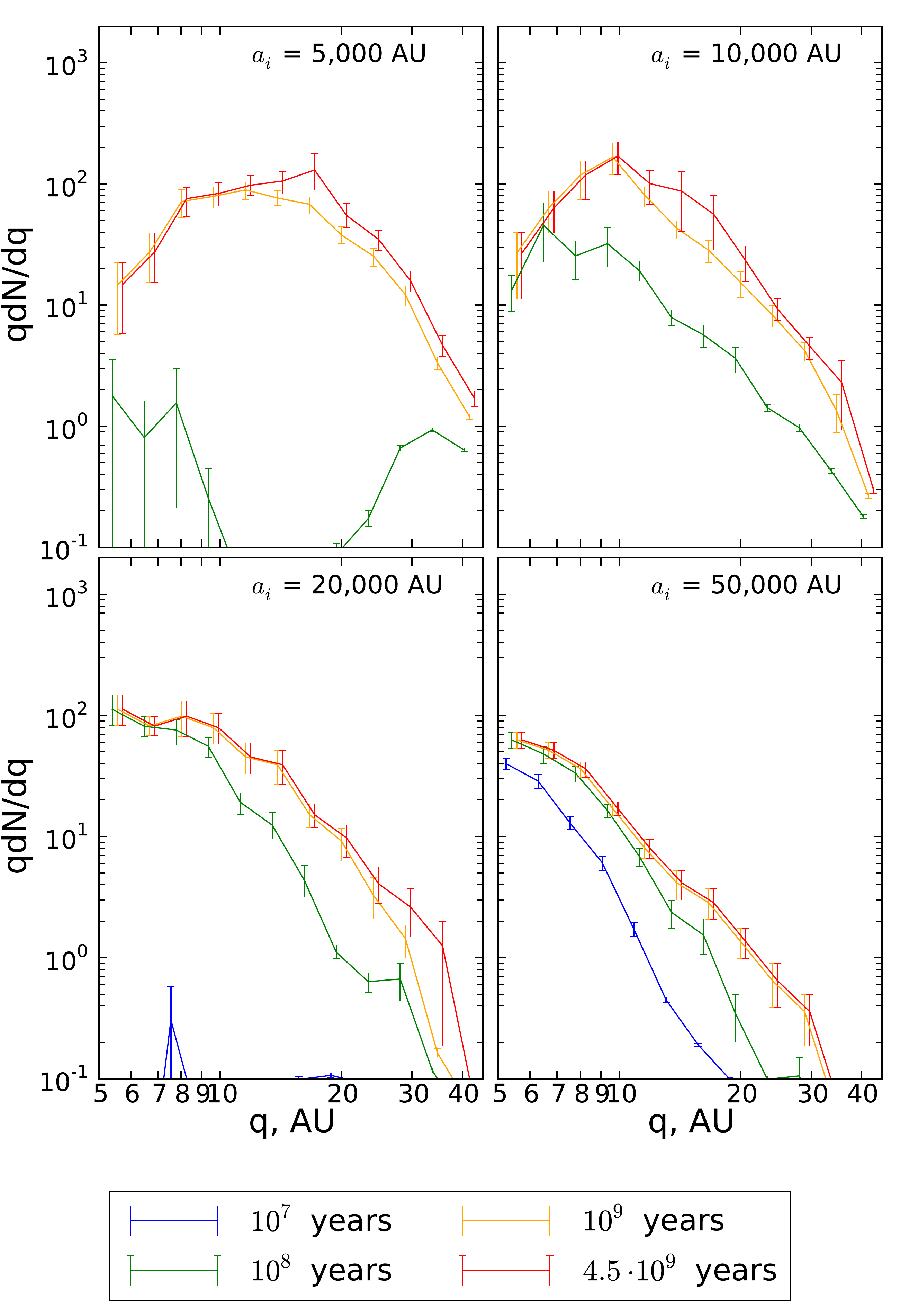}
\caption{Number of NICs with $a > 300$ AU expected to be seen per logarithmic interval in $q$ at a snapshot of time.  This assumes there are $10^{11}$ comets with $r > 1$ km at the value of $a_i$ specified in each panel.}
\label{largea}
\vspace{-.05cm}
\end{figure}  
\par
We also examined what happened if we broke up the sample into two groups depending on the current semi-major axis of the comet.  Figure \ref{largea} is identical to Figure \ref{Qvisible} except that we have only considered those comets that have semi-major axes greater than 300 AU.  The error bars are smaller, because the comets with the most appearances tend to diffuse to smaller values of $a_c$, leaving a population with less spread in number of appearances.  For this reason, this subset of comets, although smaller in number, has more power to discriminate between Oort cloud models.
 \begin{figure}
\centering
\includegraphics[width=.5\textwidth]{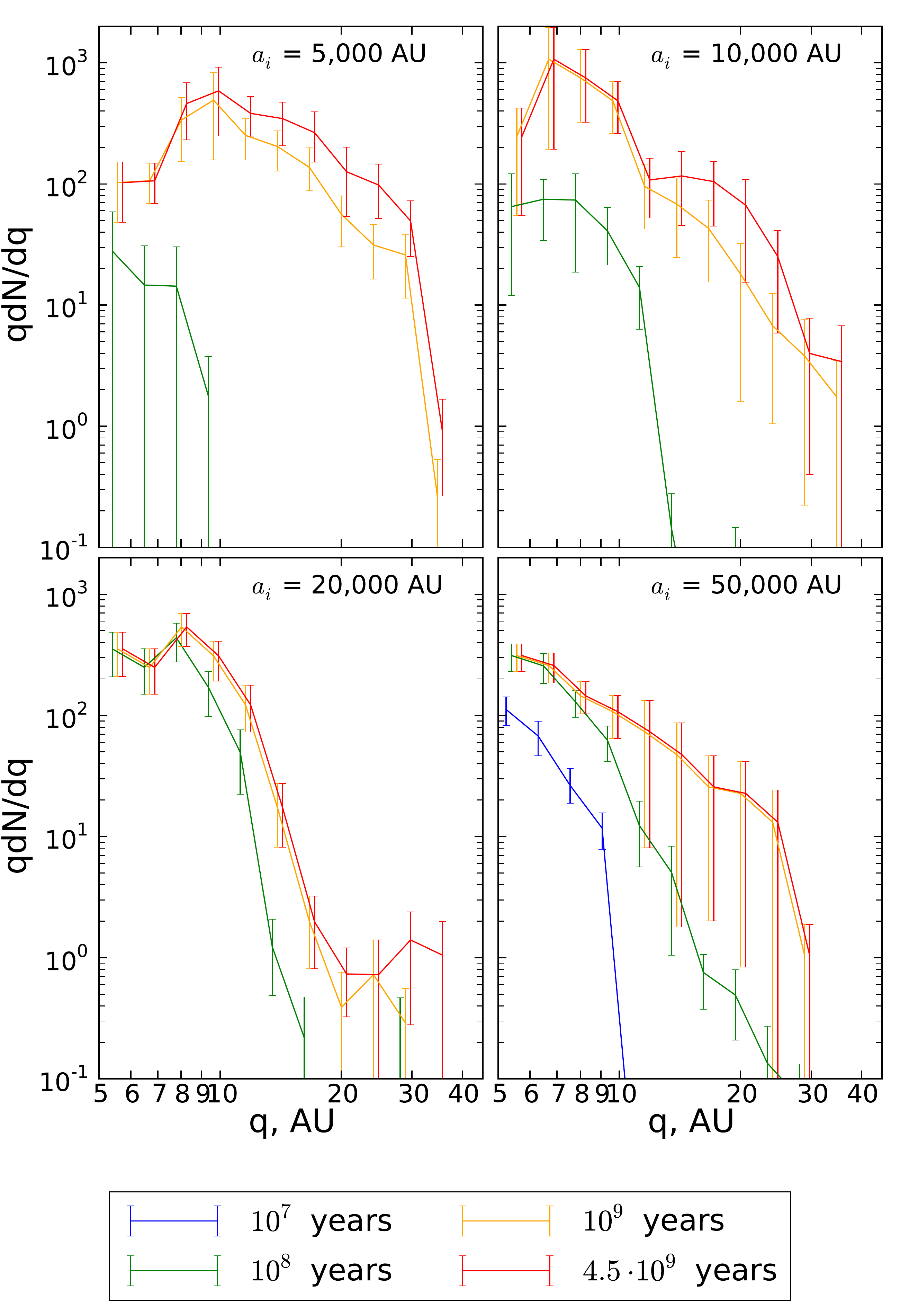}
\caption{Number of NICs with $a < 300$ AU expected to be seen per logarithmic interval in $q$ at a snapshot of time.  This assumes there are $10^{11}$ NICs with $r > 1$ km at the value of $a_i$ specified in each panel.}
\label{smalla}
\vspace{-.05cm}
\end{figure} 
In Figure \ref{smalla} we plot $qdN/dq$ for only the comets in Figure \ref{Qvisible}, but not Figure \ref{largea}, i.e., those comets whose orbits have $a_c<300$ AU.  We see that $qdN/dq$ declines more sharply with $q$ than in the whole sample of long-period comets.  This is because it is difficult for a comet to attain $a_c < 300$ AU at large perihelion, because the kicks are too small.  We also note that the overall numbers are larger by a factor of a few for comets with $a_c < 300$ AU.

\subsection{Distribution in Semi-major Axis}
\begin{figure}
\centering
\includegraphics[width=.5\textwidth]{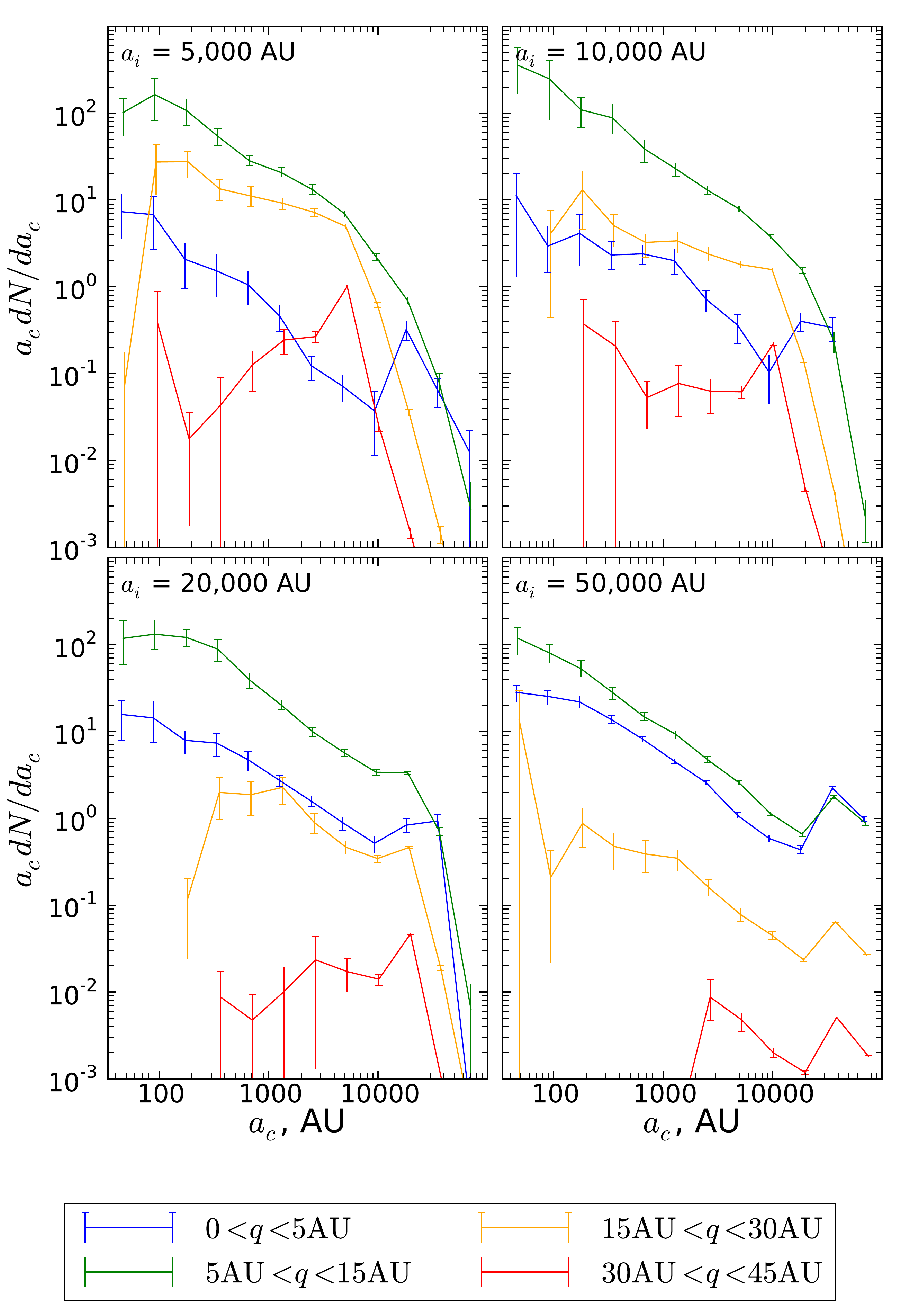}
\caption{Number of NIC appearances per logarithmic interval in semi-major axis for different values of perihelion distance (different color curves) and different values of the initial semi-major axis (different panels).  Each panel assumes that the Oort cloud contains $10^{11}$ comets with $r > 1$ km at the specified value of $a_i$.}
\label{avisible}
\vspace{-.05cm}
\end{figure} 
Figure \ref{avisible} shows the semi-major axis distribution (number of appearances per unit logarithmic interval in semi-major axis) for all the NICs in a given perihelion bin (denoted by the color of the curve) and initial semi-major axis (panel).  The error bars are generally larger for the points at small semi-major axis, implying that the statistics in these bins are dominated by a few comets. 
\par
 \begin{figure}
\centering
\includegraphics[width=.5\textwidth]{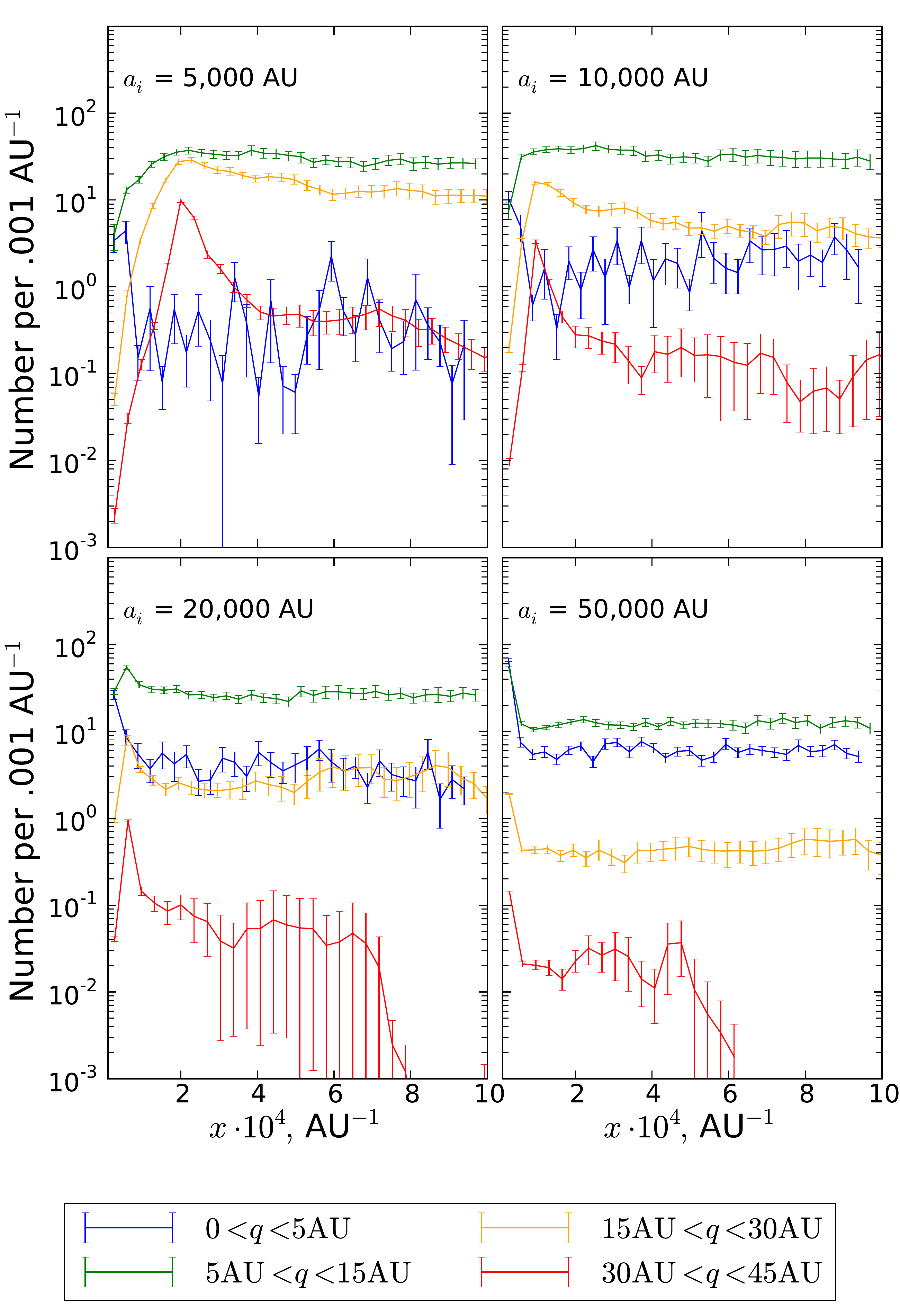}
\caption{Number of NIC appearances per linear interval in inverse semi-major axis $x = 1/a$ for different values of perihelion distance (different colors) and different values of $a_i$ (different panels).  Each panel assumes that the Oort cloud contains $10^{11}$ comets with $r > 1$ km at the specified value of $a_i$. }
\label{inversea}
\vspace{-.05cm}
\end{figure} 

We find it more illuminating to make this plot in the coordinate $x = 1/a$, which is proportional to the energy.  Doing so allows us to test the predictions of random walk models such as those in \citet{Yabushita79} and \citet{Everhart76}.  In their simplest form, one can imagine a comet starting with energy $E = -\epsilon$.  Every perihelion passage, it takes a step of size $\epsilon$ towards higher or lower energy.  It is removed if $E = 0$ (it becomes unbound), or if $E < E_{sp}$, where $E_{sp}$ is the critical energy level to be re-classified as a short-period comet.  We ignore the possibility that a short period orbit could be perturbed back to the long-period regime.  Comets are injected near $x = 0$.  In the limit that $-E_{sp}/\epsilon \gg 1$ , one can show that a steady-state distribution of comet energies normalized by the rate of perihelion passage is given by a linear equation of the form
\begin{equation}
\label{NE}
N(E) = k (E - E_{sp}),
\end{equation}
where $k$ is a constant depending on the comet injection rate and the size of the kick.  
\par 
We see some support for Equation \eqref{NE} in Figure \ref{inversea}.  We have only considered comets with $a_c > 1,\!000$ AU ($x < 0.001$ AU$^{-1}$).  This is a small enough range that we would expect the curves to be nearly flat if Equation \eqref{NE} were correct (since $E_{\rm sp}$ is much more negative than the energies shown in Figure \ref{inversea}).  The curves are generally flat for perihelion distances 5 AU $<q<$ 30 AU.  Inside 5 AU, the error bars are too large to say much.  Outside 30 AU, there is a definite trend favoring energies closer to 0, as would be expected since the diffusion rate at these perihelion distances is so slow that the steady-state energy distribution cannot be achieved. 
\par
We see evidence of a weak ``Oort spike" when $a_i = 50,\!000$ AU - an excess of comets with $x < 10^{-4}$ AU$^{-1}$ interpreted as the result of the initial conditions of cometary orbits.  This is seen only for comets with $q < 5$ AU, since for larger values of $q$, the spike is drowned out by the large numbers of comets which remain at more negative energies for many orbits.  As a quantitative measure of the observed Oort spike, \citet{Fernandez12} find that 30\% of comets discovered since the year 1900 with $q < 1.3$ AU have $x <$ (10,000 AU)$^{-1}$.  We find that we cannot come close to reproducing this unless we assume some disruption.  We applied a disruption law in which comets were removed after spending $\tau_{\rm disrupt}$ years within 5 AU of the sun.  For comets with $a_i = $ 50,000 AU and $q<5$ AU, the fraction seen with $x < 10^{-4}$ AU$^{-1}$ is \{$0.12 \pm 0.007$, $0.04 \pm 0.004$, $0.02 \pm 0.002$\}, for $\tau_{\rm disrupt}$ = \{100, 1,000, 10,000 years\}.  This confirms the well-known result that one cannot reproduce the Oort spike without some model for comet disruption or fading that removes comets after only a few appearances \citep[e.g.,][]{Wiegert99}.  
\par
We also notice a broader ``spike" of comets around the original energy for comets with $a_i = 5,\!000$ or 10,000 AU, and perihelion distances outside 30 AU.  A typical kick in energy at 30 AU might be on the order of $\Delta x = 10^{-5}$ AU$^{-1}$ \citep{Duncan87}, so comets with $a_i$ small enough to not be rapidly carried away by the tide will show a broad peak around their original energy at large perihelion distance.

\begin{figure}
\centering
\includegraphics[width=.5\textwidth]{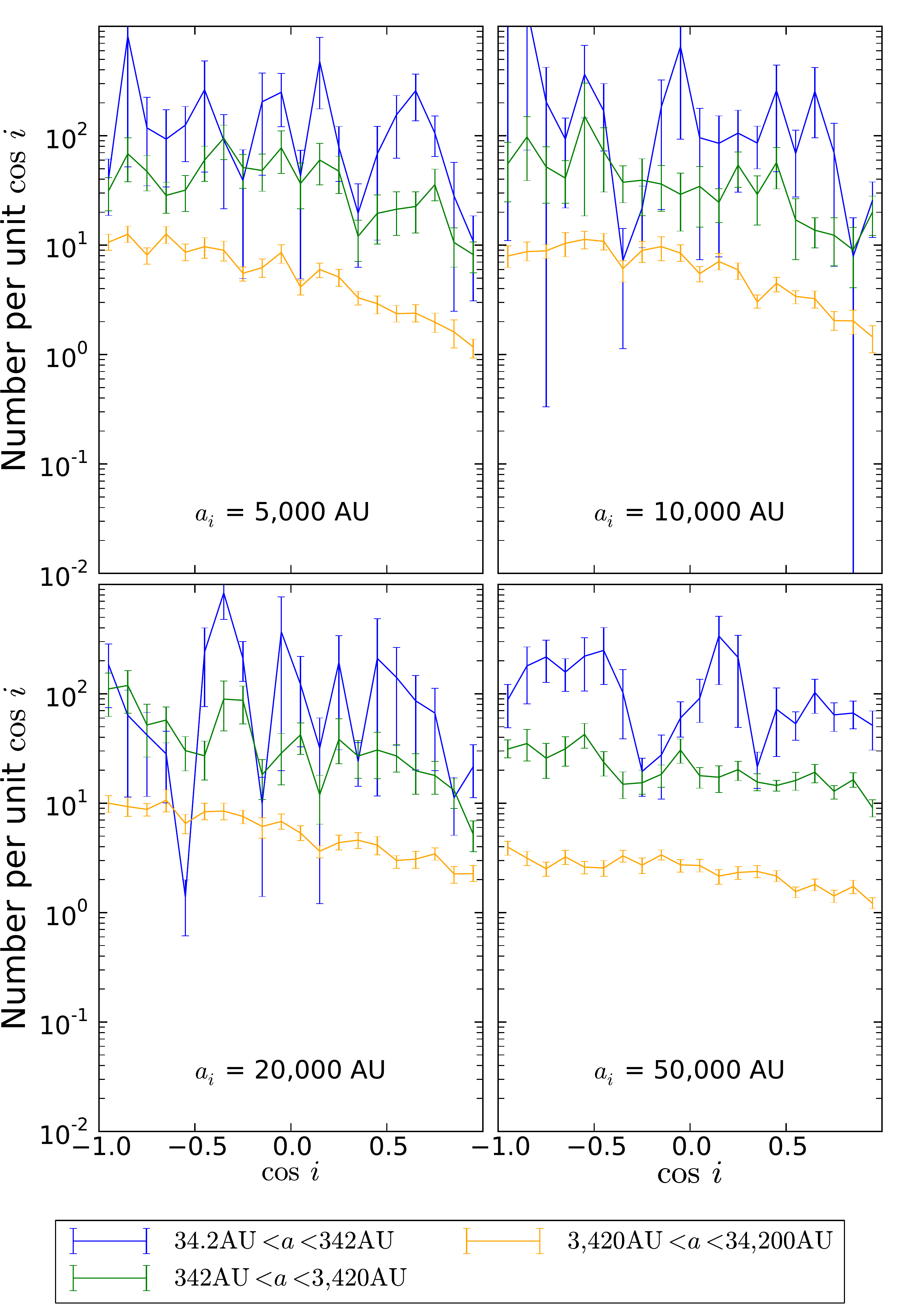}
\caption{Distribution of the cosine of the ecliptic inclination angle for different ranges of current semi-major axis (different color curves) and initial semi-major axis (different panels). }
\label{inclination}
\vspace{-.05cm}
\end{figure}

\subsection{Orbital Inclination}
Figure \ref{inclination} shows the orbital inclination distribution (in ecliptic coordinates) of the visible NICs.  The longest period NICs are preferentially retrograde.  In a random walk model, the density of comets varies inversely with the step size.  The energy kick is 2-3 times larger for a prograde comet \citep{Duncan87}, translating into an expected prograde fraction of $0.25$ to $0.33$.    
\par
We find that our error bars on the prograde fraction are large for the entire comet population, so for the following analysis we split the population into two groups depending on the semi-major axes of the comets.   Once again we find that elements of comets with large values of $a_c$ have less statistical noise.  We obtain prograde fractions among the visible comets of $0.27 \pm0.03, 0.29 \pm 0.04, 0.29 \pm 0.03$, and $0.40 \pm 0.02$ for $a_i =$ 5,000, 10,000, 20,000, and 50,000 AU respectively, for the comets with $a_c > 1,\!000$ AU.  For comets with $a_c < 1,\!000$ AU, we find prograde fractions of $0.42 \pm 0.15, 0.26 \pm 0.13, 0.32 \pm 0.09$, and $0.40 \pm 0.09$.  These data are consistent with the random walk model.
\par
 While the simulation data agree with the random walk model, they contradict the observations.  There is only a slight preference in the observational data for retrograde comets with high perihelion.  64 out of 110 comets (58\%) with period greater than 200 years and perihelion greater than 5 AU in the database at \url{http://ssd.jpl.nasa.gov/dat/ELEMENTS.COMET} are retrograde.
\subsection{Size Distribution}
A bigger telescope enables us to see rare large comets because it can search more volume.  It is impossible to say exactly what size distribution of comets to expect in the observed sample, however we can make an estimate based on extrapolation of the size distribution from \citet{Fernandez12}.  It is instructive to first consider the zero-planet model with a fixed power-law for the size distribution.  
 \begin{figure}
\centering
\includegraphics[width=.5\textwidth]{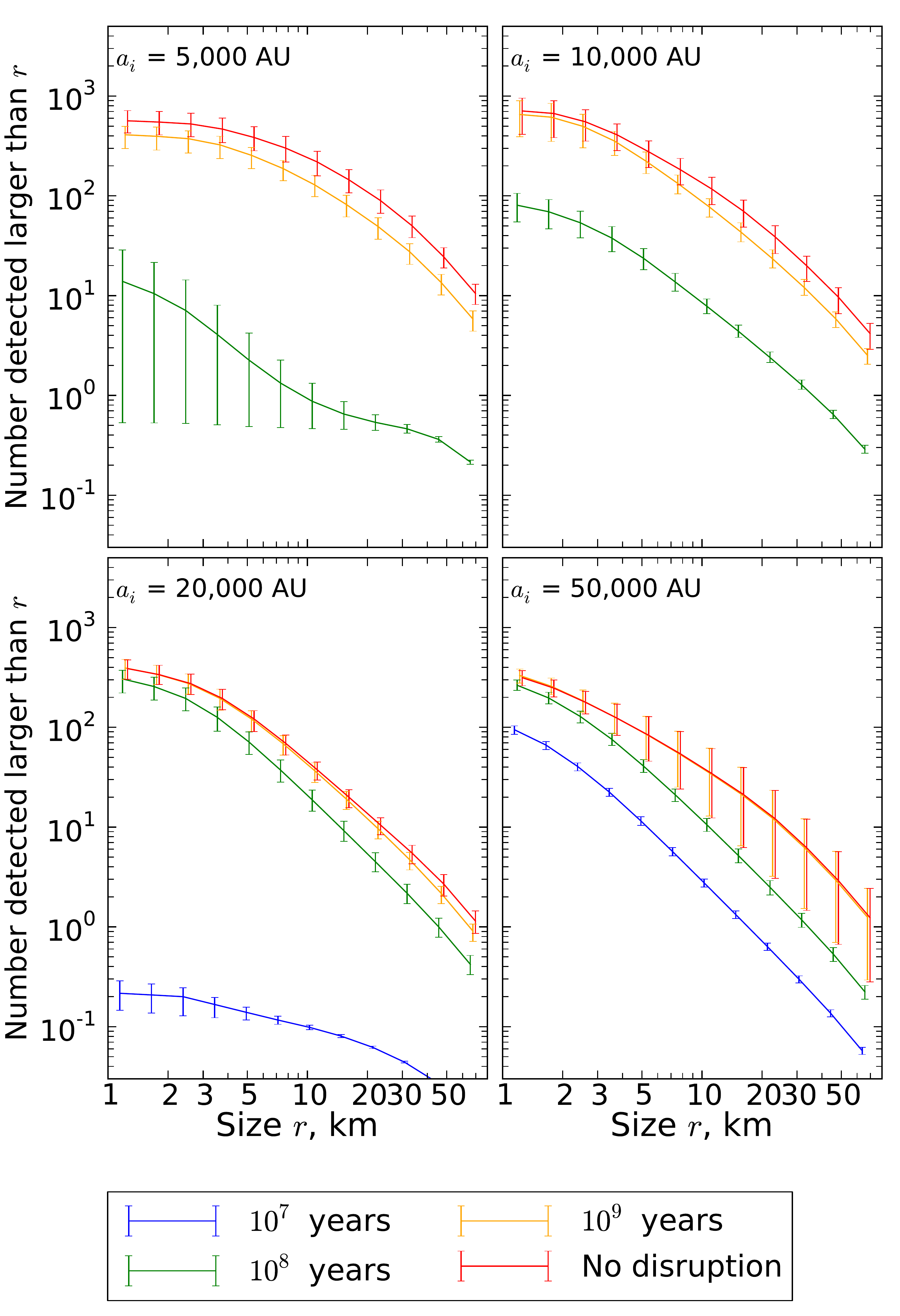}
\caption{Number of predicted detections of bodies in the range 5 AU $<R<$ 45 AU as a function of $r$ given the size distribution assumed in Equation \eqref{dNdr11}.  The blue and orange points are unmoved, and the green and red are shifted 3\% left and right respectively. }
\label{sizePlot}
\vspace{-.05cm}
\end{figure}
In Appendix \ref{sect:isovis}, we calculate the number of NICs greater than a certain size $r$ expected to be seen in the zero-planet model.  We find in Equation \eqref{eqNL} that the number visible with size greater than $r$ is given by $N = 55.2/r^{1.50}$ assuming $10^{11}$ comets greater than 1 km at 10,000 AU, but the results depend sensitively on the assumed number density and semi-major axis.   Additionally, due to the concentration at larger orbital radii, we actually expect to see objects much larger than one would infer from the zero-planet model used to derive Equation \eqref{eqNL}.  
\par
Figure \ref{sizePlot} shows the expected number of observations of NICs larger than a given size in the distance range 5 AU $<R<$ 45 AU.  This calculation assumes $10^{11}$ comets greater than 1 km at the given initial semi-major axis in the Oort cloud and slope of the size distribution given in Equation \eqref{dNdr}.  Exclusion of comets with orbital radii less than 5 AU should have a negligible effect on the counts except at the smallest sizes, as the concentration factors are lower there, and it is a negligible fraction of the visible volume for the larger comets.  Exclusion of comets with orbital radii greater than 45 AU has no effect for the size of comets that we have plotted, since none of them would be visible past 45 AU.  We see that assuming the size distribution from Equation \eqref{dNdr} holds to these sizes, we observe on the order of 10 NICs greater than 80 km in size if the majority of the NICs are coming from $a_i =$ 5,000 AU, and a few if the majority of the NICs are coming from $a_i \geq $ 10,000 AU.
\par
\section{CONCLUSIONS}
We simulated the evolution of NICs originating in the Oort cloud as they interact with the giant planets.  We used these simulations to create a catalog of simulated comet positions and velocities.  We observe different distributions of orbital elements including perihelion distance, semi-major axis and inclination depending on the semi-major axes at which the NICs originate.  Observations by LSST will therefore let us determine the absolute numbers of comets in the Oort cloud as a function of semi-major axis, and test Oort's standard model for the origin of comets.  The distribution of NICs outside the orbits of Jupiter and Saturn will provide direct evidence for the presence or absence of the hypothesized ``inner Oort cloud" corresponding to semi-major axes between $5,\!000$ and $20,\!000$ AU.
\par
One surprising result is that we expect at least tens of percent of the comets observed by LSST in the outer solar system to have been interacting with the giant planets for more than 1 Gyr.  This result makes the interpretation of the comet population detected by LSST more difficult and interesting, since the population and spatial distribution of comets in the Oort cloud almost certainly evolves on timescales of a few Gyr.
\par
We will also get a better measurement of the Oort spike --- the excess of comets in nearly parabolic orbits --- as we will be able to measure high-precision orbits in a regime where comets are likely unaffected by non-gravitational forces.  This will put constraints on models of comet fading, as well as the original semi-major axes of comets.
\par
We will furthermore be able to constrain the size distribution of Oort cloud comets out conservatively to several tens of kilometers, and perhaps even to larger bodies, depending on the size distribution and number density of comets.  Our results are based on a relatively steep slope for the size distribution of comets, $dN\propto r^{-3.76}$ (Equation \ref{dNdr}) and may substantially underestimate the total number of comets that will be discovered at large distances by LSST.
\par
We thank the referee, Ramon Brasser, for helpful and constructive comments and advice.

\bibliographystyle{apj}
\bibliography{apj-jour,LSSTPredictions}

\appendix
\numberwithin{equation}{section}
\section{MAXIMUM TORQUE ON A RADIAL ORBIT}
\label{Rtorque}
We calculate the torque on a radial orbit from the Galactic tide.  This calculation allows us to exclude comets from the Oort cloud that will definitely not enter the buffer region during the entry period $\tau_{\rm entry}$ as described in Section \ref{desc}.  The results are also used in Section \ref{sect:concentration}.  While our method is approximate, we have checked that our prediction is conservative in the sense that no comet can enter the buffer region which we predicted could not enter the buffer region.  For simplicity, we made the following approximations.  
\par
In calculating the torque, we assume that the orbit is completely radial, pointing in the direction of apocenter.  This is well-justified by the characteristic ratios of $q/a_i \approx 60/20,\!000$.  We ignore the components of the Galactic tidal field in the plane of the Galaxy, as they are smaller than the out-of-plane ($z$) component by about a factor of 10. 
\par
\citet{Heisler86} give the tidal force (per unit mass) as
\begin{equation}
\label{fz}
F_z = \left[-4\pi G \rho_0 + 2(B^2-A^2) \right]z_{\rm gal} \hat z_{\rm gal},
\end{equation}
where $\rho_0 = 0.09\, {\rm M_\odot}\, {\rm pc}^{-3}$, and the Oort $A$ and $B$ constants are taken to be 14.6 km s$^{-1}$ kpc$^{-1}$ and $-12.4$ km s$^{-1}$ kpc$^{-1}$ respectively \citep{BinneyAndTremaine}. This leads to an instantaneous torque given by 
\begin{equation}
\label{gamma}
\Gamma = \left[-4\pi G \rho_0 + 2(B^2-A^2) \right] \cdot \frac{R^2\sin{2b}}{2},
\end{equation}
where $b$ is the Galactic latitude of the comet (which is constant in the radial orbit approximation).  The orbit averaged torque is calculated by noting that $\langle R^2 \rangle = 5a^2/2$.
\par

\section{COMETS VISIBLE ASSUMING A UNIFORM DISTRIBUTION OF ORBITAL ELEMENTS ON AN ENERGY SURFACE AND NO PLANETS}
\label{sect:isovis}
In this appendix, we calculate the number of comets observable at a snapshot in time as a function of heliocentric radius in the zero-planet model.  We approximate comet orbits as parabolic in the observation region.  If we have a population of comets with perihelion $q$, of which $s(R)$ pass perihelion per year that are large enough to be visible at $R$, then the density $N_{\rm vis}(R) dR$ of comets visible in a radial interval at radius $R>q$, is
\begin{equation}
\label{dNdR}
N_{\rm vis}(R)  = \frac{2s(R)}{v_R} ,
\end{equation} 
where $v_R$ is the radial velocity (Equation \eqref{vr}).  Using Equations \eqref{dNdr} and \eqref{rR}, we find that in the zero-planet model,
\begin{equation}
\label{sR}
s(R) =  \frac{-c_0}{\beta}\mathlarger{e^{-\gamma \alpha/\beta}\int_{r_1R^2}^\infty r^{\gamma/\beta -1}}dr = \frac{c_0}{\gamma}\mathlarger{e^{-\gamma \alpha/\beta} r_1^{\gamma/\beta} R^{2 \gamma/\beta}}
\end{equation}
where $r_1= 0.0361 \cdot 10^{\mathlarger{\frac{24.5-m_{\rm lim}}{5}}}$ is the size (in km) of a body that is marginally detectable at 1 AU in our model (see Equation \eqref{rR}).  Therefore, using Equations \eqref{vr}, \eqref{dNdR} and \eqref{sR} we can write the density of comets as an integral over $q$:
\begin{align}
\label{noPlanetNorm}
& N_{\rm vis}(R) =  2\frac{c_0}{\gamma}\mathlarger{e^{-\gamma \alpha/\beta} r_1^{\gamma/\beta} R^{2 \gamma/\beta}} \int_0^R \frac{\sqrt{R}dq}{ 2\pi \sqrt{2(1-\frac{q}{R})}} =
\nonumber\\
& \frac{\sqrt{2}c_0}{\pi \gamma}\mathlarger{e^{-\gamma \alpha/\beta} r_1^{\gamma/\beta} R^{2 \gamma/\beta + 3/2}} = 5.0\cdot 10^{0.55(m_{\rm lim} - 24.5)} \left(\frac{R}{5}\right)^{-4.02}.
\end{align}
\par
We also calculate how many comets we see as a function of perihelion given a magnitude cut at a given snapshot in time.   We find that
\begin{equation}
N_{\rm vis}(q) =\frac{-c_0}{\beta}\mathlarger{e^{\frac{-\gamma \alpha}{\beta}}\int_{r_1q^2}^\infty r^{\frac{\gamma}{\beta} -1}} \tau_{\rm vis} (r, q) dr,
\end{equation}
where $\tau_{\rm vis}(r, q)$ is the amount of time that a comet with radius $r$ in kilometers and perihelion $q$ in AU is visible during one perihelion passage.  From Equation \eqref{vr}, we see that
\begin{equation}
\label{tauvis}
\tau_{\rm vis}(r, q) = 2 \int_q^{R_{\rm max}} dR(v_R)^{-1} = \frac{\sqrt{2}}{3 \pi} (2q + R_{\rm max})\sqrt{R_{\rm max} - q} ,
\end{equation}
where $R_{\rm max} = \sqrt{r/r_1}$ is the maximum heliocentric distance at which the comet is visible.
Putting this together, we find that 
\begin{equation}
\label{noPlanetQ}
N_{\rm vis}(q) = c_0 \mathlarger{e^{-\gamma \alpha/\beta}} \left[\frac{ r_1^{\frac{\gamma}{\beta}} \Gamma(\frac{-3}{2} - \frac{2 \gamma}{\beta})}{\sqrt{2 \pi} \gamma \Gamma(-1 - \frac{2 \gamma}{\beta})}\right] \mathlarger{q^{3/2 + 2 \gamma/\beta}} = 2.3\cdot 10^{0.55(m_{\rm lim} - 24.5)}.
\left(\frac{q}{5}\right)^{-4.02}.
\end{equation}
\par
Finally, we ask how many comets we expect to see above a certain size in the zero-planet model.  In this calculation, we take the size distribution in Equation \eqref{dNdr}, but normalize the counts to those expected if there were to $10^{11}$ comets with semi-major axis $a_i$ and radii greater than 1 km.  A cloud of $N$ comets with orbital elements uniformly distributed on a surface of constant energy in phase space will have $2/a_i^{5/2}$ comets passing perihelion per year per AU perihelion, for $q \ll a_i$.  Therefore, the correct normalization is
\begin{equation}
\label{dNdr11}
dN_{\rm peri}/dr =-\frac{2N}{a_i^{\frac{5}{2}}}\frac{\beta}{\gamma}r^{\frac{\gamma}{\beta} - 1}  =7.25\cdot r^{-3.76} \frac{N}{10^{11}} \left(\frac{10^4 \; {\rm AU}}{a_i}\right)^{\frac{5}{2}}.
\end{equation}
\par

In this model, we can expect to see $N_L(r)$ comets larger than $r$ where $N_L$ is given by 
\begin{eqnarray}
\label{eqNL}
&N_L(r) = \mathlarger{\int_{r' = r}^\infty  -\frac{2N}{a_i^{\frac{5}{2}}}\frac{\beta}{\gamma}r'^{\frac{\gamma}{\beta}-1}\int_0^{R_{\rm max}} \tau_{\rm vis}(r', q)}dqdr' = \nonumber\\ 
&  \mathlarger{\mathlarger{\frac{16 \sqrt{2} \beta^2 N r^{\frac{5}{4} + \frac{\gamma}{\beta}}}{5 \pi \gamma a_i^{\frac{5}{2}} r_1^{\frac{5}{4}}(5 \beta + 4 \gamma)} = \frac{55.3 }{r^{1.50}} \left(\frac{10^4 \; {\rm AU}}{a_i}\right)^{\frac{5}{2}}\cdot 10^{0.25(m_{\rm lim} - 24.5)}}}.
\end{eqnarray}
Setting this equal to $1$, and solving for $r$, we find that we expect to see one comet larger than 14.4 km if $a_i$ = 10,000 AU, but this estimate is clearly quite sensitive to the assumed number of comets and that value of $a_i$.
\end{document}